\documentclass[12pt]{article}
\usepackage{latexsym}
\usepackage{epsf}

\font\mybb=msbm10 at 12pt
\def\bb#1{\hbox{\mybb#1}}
\def\Z {\bb{Z}}

\makeatletter
\@addtoreset{equation}{section}
\makeatother

\def\unit{\hbox to 3.3pt{\hskip1.3pt \vrule height 7pt width .4pt \hskip.7pt
\vrule height 7.85pt width .4pt \kern-2.4pt
\hrulefill \kern-3pt
\raise 4pt\hbox{\char'40}}}

\def\be{\begin{equation}}
\def\ee{\end{equation}}
\def\bea{\begin{eqnarray}}
\def\eea{\end{eqnarray}}

\def\part{\partial}

 \def\makeatletter{\catcode`\@=11}
\makeatletter
\def\mathbox#1{\hbox{$\m@th#1$}}
\def\math@ccstyles#1#2#3#4#5#6#7{{\leavevmode
      \setbox0\mathbox{#6#7}%
      \setbox2\mathbox{#4#5}%
      \dimen@ #3%
      \baselineskip\z@\lineskiplimit#1\lineskip\z@
      \vbox{\ialign{##\crcr
             \hfil \kern #2\box2 \hfil\crcr
             \noalign{\kern\dimen@}%
             \hfil\box0\hfil\crcr}}}}
\def\mathaccstyles{\math@ccstyles\maxdimen}
\def\maththroughstyles{\math@ccstyles{-\maxdimen}}
\def\unitmatrixDT%
 {\maththroughstyles{.45\ht0}\z@\displaystyle {\mathchar"006C}\displaystyle 1}

\begin{document}
\begin{flushright}
\footnotesize
\footnotesize
UG-16/99\\
{\bf hep-th/9908121}\\
August, $1999$
\normalsize
\end{flushright}

\begin{center}

\vspace{1.5cm}
{\Large {\bf Stable D8-branes and Tachyon Condensation}}\\
\vspace{.4cm}
{\Large {\bf in Type 0 Open String Theory}}

\vspace{1.5cm}


{\bf Eduardo Eyras}

\vspace{.4cm}

{
{\it Institute for Theoretical Physics\\
University of Groningen \\
Nijenborgh 4, 9747 AG Groningen, The Netherlands}\\
{\tt E.A.Eyras@phys.rug.nl}
}

\vspace{.5cm}
\vspace{.3cm}
\vspace{.8cm}


{\bf Abstract}
\end{center}

\begin{quotation}
\small

We consider non-BPS D8 (and D7) branes in type 0 open string theory and
describe under which circumstances these branes are stable.
We find stable non-BPS D7 and D8 
in type 0 with and without D9-branes in the background.
By extending the descent relations between D-branes to
type 0 theories, the non-BPS D8-brane is considered as the 
result of a tachyon condensation of a D9 anti-D9 pair in type 0.
We study the condensation of the open string tachyons
in type 0 with generic gauge groups giving rise to different
configurations involving non-BPS D8-branes and discuss the stability 
in each case. The results agree with the topological analysis of the
vacuum manifold of the tachyon potential for each case.

\end{quotation}





\newpage
\section*{Introduction}

In the past year there has been a tremendous increase 
in our knowledge about non-BPS states and non-BPS branes
in string theory \cite{Sen-0,Sen-review,Lerda-review}. 
We have learned that only under certain 
circumstances these branes can be stable. This fact has been used in order
to extend the duality checks to the non-BPS part of spectrum of the 
string theories 
\cite{Sen-review,Lerda-review,Sen-1,Bergman-Gaberdiel-4,Sen-2,
Bergman-Gaberdiel-3}.
These developments find a natural implementation in the context of 
non-supersymmetric strings, where the lack of spacetime supersymmetry
forces one to find other stability conditions for the solitons of the theory, 
which allow one to establish duality relations to some
extent \cite{Bergman-Gaberdiel-1,Gates-Rodgers,Blum-Dienes,
Harvey-1,Bergman-Gaberdiel-2,Michishita}.
Moreover, there are indications that non-supersymmetric strings
might find their place in the 
unification picture of M-theory \cite{Bergman-Gaberdiel-2}.
It is then natural to try to check these dualities by means of
stable non-BPS states in the non-supersymmetric theory\footnote{Since we do 
not have spacetime supersymmetry, non-BPS means in this context that there is
no fixed relation between the mass and the charge.}.
Thus it is interesting to find stable non-BPS 
states and in particular stable non-BPS D-branes 
in non-supersymmetric string theories.
Progress in this context has been reported in \cite{Michishita},
where a stable D-particle was constructed in 
type 0 open string theory with gauge group $SO(32) \times SO(32)$.

In this paper we describe the non-BPS D7 and D8 branes of type 0
open string theory and in particular, we discuss under which circumstances they
are stable. We use the fact that in type 0 theory the background
of D9-branes is not completely determined by the tadpole
cancellation, and study the different possibilities.
This makes possible to find that, 
although non-BPS D7 and D8 branes are not stable in
type I \cite{Lerda-nonBPS}, they can be stable in 
type 0.
Non-BPS D$p$-branes appear in general as a kink solution in the 
tachyon potential of a D$(p+1)$-brane anti-D$(p+1)$-brane pair
\cite{Sen-2,Sen-3}. 
Similarly we consider these D8-branes in type 0 as the
result of tachyon condensation of the D9 anti-D9 pairs
in type 0 open string theory. More generally, we analyze
the mechanism of tachyon condensation of the D9 anti-D9 pairs giving
rise to non-BPS D8-branes for generic type 0 backgrounds
with symmetries
$SO(n) \times SO(m) \times SO(n) \times SO(m)$,
$SO(n) \times SO(n) \times \Z_2$, $SO(n) \times SO(n)$ and $\Z_2 \times \Z_2$.
We study the stability in terms of the
resulting brane configuration and using topological arguments; 
and show that both results agree.

This article is organized as follows. In section 1
we describe the type 0 open string theory first considered in 
\cite{Bianchi-Sagnotti-1}. 
We briefly review the tadpole cancellation with 
D9-branes, which can give rise to groups
of the form $SO(N) \times SO(32-N) \times SO(N) \times SO(32-N)$.
We also show how to obtain the same group structure from an orbifold
of type I.
In section 2 we show that a non-BPS D8$_+$ (D8$_-$) brane is stable if
(1) there are only D9$_-$ and anti-D9$_-$ (D9$_+$  and anti-D9$_+$) 
branes in the background
or when (2) there are no D9-branes at all in the background.
These results are  completely analogous for the non-BPS D7-brane.
Finally, in section 3 we describe the
condensation of the open string tachyons
between the D9-branes and anti-D9-branes in type 0 open string theory
for different generic situations.

\section{Type 0 open string theory}

Type 0B (0A) theory is obtained as an orbifold of type IIB (IIA) by
the spacetime fermion number operator $(-1)^{F_s}$
\cite{Dixon-Harvey}.
The spectrum of the type 0B (0A) theory and of most of their open descendants
contains double the number of R-R fields with respect to the type II case.
We denote these two fields as:
\be
(R+,R\pm) \Rightarrow C^{(p+1)} \, ,\qquad 
(R-,R\mp) \Rightarrow {\bar C}^{(p+1)}
\, ,
\label{RR-fields-1}
\ee
where the upper signs\footnote{These signs refer to the
worldsheet fermion numbers of the left and right moving sectors,
i.e. the eigenvalues of $(-1)^{F_L}$ and $(-1)^{F_R}$, respectively.} 
correspond to the type 0B with $p$ odd, and
the lower signs correspond to type 0A with $p$ even.
We consider the combinations \cite{Klebanov-Tseytlin}:
\be
C^{(p+1)}_{\pm} = {1 \over \sqrt{2}} ( C^{(p+1)} \pm {\bar C}^{(p+1)} ) \, .
\ee
Consequently, we have two types of D$p$-branes for each $p$, that we denote
as D$p_+$ and D$p_-$, which carry one unit of charge 
under $C^{(p+1)}_+$ and $C^{(p+1)}_-$, respectively. The relation between these
charges and the charges of the R-R fields in (\ref{RR-fields-1}) is given by
\be
q_\pm = {1 \over 2} (q \pm {\bar q}) \, .
\ee
Type 0 theory can be obtained as an orientifold of type 0B by $\Omega$
\cite{Bianchi-Sagnotti-1}, or as an 
orbifold by $(-1)^{F_s}$ of type I \cite{Bergman-Gaberdiel-1}.
The projection of type I by 
$(-1)^{F_s}$ eliminates all the fermions from the spectrum.
The (massless) untwisted sector is given by the metric, the
dilaton, and a R-R 2-form $C^{(2)}$, in the sector $(R+,R+)$.
The twisted sector introduces a tachyon $(NS-,NS-)$
and a second R-R 2-form ${\bar C}^{(2)}$ in the sector $(R-,R-)$.
The orbifold  $(-1)^{F_s}$
has the same effect as a diagonal GSO projection. Accordingly,
the contribution of the Klein bottle to the one-loop vacuum amplitude
only gives rise to a NS-NS massless tadpole in the tree channel,
hence the Orientifold fixed plane of type 0 carries no R-R 
charge\footnote{The Klein-Bottle contribution in type 0 is the double of the
type I case, so one could consider one Orientifold and one
anti-Orientifold \cite{Bergman-Gaberdiel-Lifschytz} 
in type 0 \cite{Kachru-Kumar-Silverstein}. However, 
since they carry no R-R charge, they cannot be distinguished in type 0.}.
The NS-NS tadpole does not render the theory necessarily inconsistent,
hence in principle we do not need to introduce D9-branes, 
so we could do without an open-string sector in the theory. 
The NS-NS can in fact be
removed by a Fischler-Susskind mechanism \cite{F-S}, which introduces
a spacetime dependent coupling.

\vskip 12pt
\noindent{\bf Tadpole Cancellation in Type 0}
\vskip 6pt

The NS-NS tadpole of type 0 can also be canceled by adding
D9-branes, which introduce an open sector in the theory.
Since only the cylinder 
diagram contributes to a R-R massless tadpole, 
this must be
canceled by the D9-branes themselves, hence we have to take an equal number
of D9-branes and anti D9-branes. 
The cancellation of the NS-NS tadpole tells us that the
number of branes to be introduced is $N=32$:
\be
 {\cal A}^{NSNS} = - V_{10} 
\int_0^\infty {d \ell \over 2} 
(8\pi^2 \alpha^\prime) 
\left( {1 \over 2} (2N)^2 {\rm e}^{2\pi \ell} + 16 (N-32)^2 \right) \, .
\ee
The exponentially divergent term is a tachyon tadpole, which is expected in a 
theory with a tachyon in the closed string spectrum.
Thus type 0 string theory without tadpoles
can have a gauge group $SO(32) \times SO(32)$.

On the other hand, we know that
type 0B contains two types of D$p$-branes for each $p$
odd. This implies that we also have two types of D9-branes in the type 0
theory \cite{Bergman-Gaberdiel-1,Bergman-Gaberdiel-2}. 
Thus regarding the type 0 theory as a type 0B orientifold, 
there are different possible combinations of D9-branes
which cancel the tadpoles. The generic case is a system of 
$N$ D9$_+$ branes and $32-N$ D9$_-$ branes, with their respective antibranes.
This configuration originates the gauge group
$SO(N) \times SO(32-N) \times SO(N) \times SO(32-N)$, and
the R-R and NS-NS massless tadpoles vanish.
Moreover, the tachyon tadpole is slightly changed:
\be
 {\cal A}^{NSNS} = - V_{10} 
\int_0^\infty {d \ell} \, 
(8\pi^2 \alpha^\prime) 
\left(32 - 2N \right)^2 {\rm e}^{2\pi \ell} \, ,
\label{tachyon-tadpole}
\ee
and vanishes for $N=16$. This is the situation where we have
16 D9$_+$ and 16 D9$_-$ branes and their respective antibranes,
which is equivalent to 16 bound states of the form D9$_{\pm}$
plus the corresponding antibranes. The fact that the
tachyon tadpole vanishes\footnote{Although the closed string
tachyon does not appear in the cylinder amplitude for $N=16$, it does appear
in the torus amplitude.} in this specific configuration
is consistent with the 
fact that the potential between the D9$_\pm$ bound states 
is proportional to the potential between type I D9-branes.
One concludes that from the point of view of a type 0B orientifold,
type 0 without tadpoles can have in general
$SO(N) \times SO(32-N) \times SO(N) \times SO(32-N)$ symmetry.

\newpage
\noindent{\bf Type 0 as an Orbifold of Type I}
\vskip 6pt

The generic configuration of D9-branes with gauge group
$SO(N) \times SO(32-N) \times SO(N) \times SO(32-N)$ 
can also be made consistent with the 
orbifold  construction of type 0 from type I\footnote{This extents
the result of \cite{Bergman-Gaberdiel-1}, where only the
$SO(32) \times SO(32)$ case was obtained as an orbifold of type I.}.
Consider the gauge transformation
\be
{\cal I} = \left( \begin{array}{cc} -\unitmatrixDT_N & 0 \\
					0    & \unitmatrixDT_{32-N} 
\end{array} \right) \, ,
\label{cal-I}
\ee
which acts on the Chan-Paton factors of the open string states as
\be
| \dots {\cal i} \otimes \Lambda \longrightarrow
| \dots {\cal i} \otimes {\cal I} \Lambda {\cal I}^{-1} \, .
\ee 
The orbifold\footnote{This type of projection was used
previously in \cite{Klebanov-Tseytlin-2} to connect the D3-brane of type IIB 
with the D3$_\pm$ bound state of type 0B.}
of type I by $(-1)^{F_s} \cdot {\cal I}$
then yields the type 0 theory with gauge group
$SO(N) \times SO(32-N) \times SO(N) \times SO(32-N)$.
This construction goes schematically as follows.
The group element ${\cal I}$ makes a natural division of
the 32 D9-branes of type I into two sets, one with
$N$ branes and the other one with $32-N$ branes. Accordingly, the
Chan-Paton factors for open strings can be divided as
\be
\Lambda = \left( \begin{array}{cc} A & B \\ C & D \end{array} \right) \, ,
\ee
where $A$ is an $N\times N$ matrix associated to the $N$ D9-branes
of the first set, and $D$ is a
$(32-N) \times (32-N)$ matrix associated to the other $32 - N$ D9-branes.
$B$ and $C$ are the Chan-Paton factors for the open strings stretching
from one of the $N$ branes of the first set and 
another one from the second set with
$32-N$ branes. Orbifolding by $(-1)^{F_s} \cdot {\cal I}$
we eliminate the fermions in the open strings which begin 
and end on branes of the same set, and the bosons of the open strings
with each end in a brane of either set.
The gauge group $SO(32)$ is thus broken down to
$SO(N)\times SO(32-N)$ and
we are left with massless fermions in the $({\bf N},{\bf 32-N})$
representation. This implies that the 32 D9-branes become
$N$ D9-branes of one type and $32-N$ D9-branes of the other one, 
either D9$_+$ or D9$_-$.

\begin{figure}[!ht]
\begin{center}
\leavevmode
\epsfxsize= 7cm
\epsfysize= 4cm
\epsffile{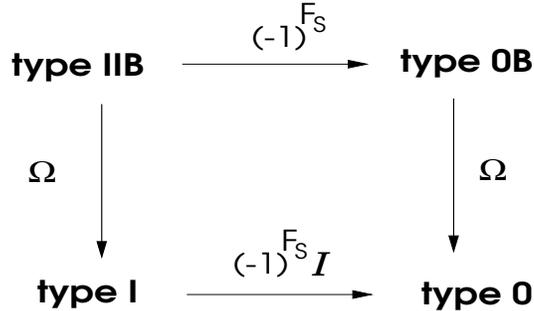}
\caption{\small {\bf Type 0 theory.} This graphic shows how to construct
type 0 theory from type 0B and type I, with a
gauge group of the form $SO(N) \times SO(32-N) \times SO(N) \times SO(32-N)$.
The operator ${\cal I}$ determines $N$ in the construction from type I.
This choice corresponds 
in the orientifold case to a certain gauge freedom we have in order to choose
the types of branes canceling the tadpole.}
\label{dibujo1}
\end{center}
\end{figure}

The anti D9-branes can be considered as coming from the twisted sector
of the orbifold $(-1)^{F_s} \cdot {\cal I}$.
Implementing the orbifold projection on this twisted sector we find
$N$ anti D9-branes of one type and $32-N$ of the other type.
Moreover, tadpole 
cancellation\footnote{One might think that there is an ambiguity in the
choice of type of brane in the twisted sector. However, in a configuration
with $N$ D9$_+$ branes, $32-N$ D9$_-$ branes, $32-N$ anti D9$_+$ branes and
$N$ anti D9$_-$ branes, the tachyon tadpole vanishes for any $N$, but the R-R
tadpole would only vanish for $N=16$.}  
imposes that the $N$ anti D9-branes must be
of the same type as the $N$ D9-branes of the untwisted sector, and similarly 
for the other $32-N$ D9-branes.
The operation ${\cal I}$ does not act on the closed string spectrum. Thus 
after orbifolding type I by $(-1)^{F_s} \cdot {\cal I}$ 
we obtain the same closed spectrum as before.
Finally, we obtain the type 0 theory with the group 
$SO(N) \times SO(32-N) \times SO(N) \times SO(32-N)$, containing
both D9$_+$ and D9$_-$ branes.

Finally, we would like to emphasize that the introduction of the D9-branes is 
completely arbitrary, 
since there is no R-R tadpole to cancel and the NS-NS tadpole is
harmless and can be treated with a Fischler-Susskind mechanism
\cite{F-S}. Accordingly, we can consider a
generic type 0 theory with
$n$ D9$_+$ branes and $m$ D9$_-$ branes, and the corresponding
antibranes. The gauge group is
$SO(n) \times SO(m) \times SO(n) \times SO(m)$ and there is a NS-NS
tadpole unless $n+m = 32$.
In particular, we can consider type 0
with no D9-branes at all, hence the open sector would be absent.
From now on we will consider these more generic backgrounds for type 0. 
The transition from one case to the other with less branes
can be seen as produced by the annihilation
of D9 anti-D9 pairs into the vacuum, where the open string
tachyon condenses restoring the
vacuum. In section 3 we will propose an alternative tachyon condensation
for which the D9 anti-D9 pair gives rise to a non-BPS D8-brane.
We proceed now with the description of this non-BPS D8-brane.

\section{Stable D8-brane in Type 0}


Using the descent relations between D-branes in type II theories
\cite{Sen-5,Sen-review}, one can 
construct a non-BPS D$(2p)$-brane in type IIB theory. 
This can be obtained either from a D$(2p+1)$ anti-D$(2p+1)$ pair in type IIB,
or from a D$(2p)$ anti-D$(2p)$ pair of type IIA.
We can extend these
relations to type 0A/0B theories.
For instance, in a D$(2p+1)_+$ anti-D$(2p+1)_+$ pair in type 0B
there is a non-BPS D$(2p)_+$ brane associated to the tachyonic kink
solution. This construction goes through like in type II, except for the
R-sector which is absent.
This non-BPS D$(2p)_+$ brane can also be constructed from
a D$(2p)_+$ anti-D$(2p)_+$ pair of type 0A.
In order to obtain this, we argue that the
orbifold projection of type 0A (0B)  by
$(-1)^{F^s_L}$ yields type 0B (0A), where
$(-1)^{F^s_L}$ is the spacetime fermion-number of the left-movers and
acts with a minus sign on the R-R sectors. 
In this way we can start with
a D8$_+$ anti-D8$_+$ pair in type 0A and by projecting with
$(-1)^{F^s_L}$ we obtain a non-BPS D8$_+$-brane in type 0B, or else,
we can construct the non-BPS D8$_+$ brane as a kink solution  
of the tachyon potential in a D9$_+$ anti-D9$_+$ pair in type 0B.
The low-energy field content on this non-BPS D8$_+$ brane in type 0B
is the same as for the
non-BPS D8-brane of type IIB but without fermions: an $U(1)$ vector,
a transversal scalar and a tachyon. This can also be obtained from
the non-BPS D8-brane in type IIB by projecting with the
orbifold $(-1)^{F_s}$.

Notice that we can choose between two types of branes, 
either D$p_+$ or D$p_-$.
We can distinguish
a non-BPS D8$_+$ from a non-BPS D8$_-$ in type 0B
at low energies by the sign of the coupling to the closed string tachyon. 
In order to see 
this we compare the cylinder amplitude between two non-BPS D8$_+$
branes
\be
{\cal A}_{{\rm D8}_+ {\rm -} {\rm D8}_+}
= V_9 \int_0^\infty {dt \over 2t} (8\pi^2 \alpha^\prime t)^{- {9 \over 2}}
\, {\rm e}^{ - {Y^2 t \over 2\pi \alpha^\prime}} \, 
{ f_3^8 ({\rm e}^{-\pi t}) \over f_1^8 ({\rm e}^{-\pi t})} \, ,
\ee
with the amplitude between a non-BPS D8$_+$ and a non-BPS D8$_-$:
\be
{\cal A}_{{\rm D8}_+ {\rm -} {\rm D8}_-}
= - V_9 \int_0^\infty {dt \over 2t} (8\pi^2 \alpha^\prime t)^{- {9 \over 2}}
\, {\rm e}^{ - {Y^2 t \over 2\pi \alpha^\prime}} \, 
{f_2^8 ({\rm e}^{-\pi t}) \over f_1^8 ({\rm e}^{-\pi t})} \, .
\label{amplitude:D8+/D8-}
\ee
We factorize the cylinder in the closed string channel, expanding
for small $t$. We find
\be
{ f_3^8 ({\rm e}^{-\pi t}) \over f_1^8 ({\rm e}^{-\pi t})} 
\sim t^4 \left( {\rm e}^{\pi / t} +8 \right) \, ,
\qquad
- \, { f_2^8 ({\rm e}^{-\pi t}) \over f_1^8 ({\rm e}^{-\pi t})} 
\sim t^4 \left( - {\rm e}^{\pi / t} +8 \right) \, .
\ee
Thus we can conclude that 
they couple with a different sign to the closed string tachyon.
Notice also that the bound state D8$_\pm$ has a cylinder amplitude
proportional to that of a non-BPS D8-brane in type IIB.

In order to obtain the D8$_+$ brane of type 0 theory we must consider the
projection by $\Omega$ in the D8$_+$ brane of type 0B.
From \cite{Lerda-nonBPS} we know that the action of $\Omega$
on the 8-8 strings is the following. 
For the $(-1)^F$ NS even part we find
\be
\Omega \, \psi^i_{-1/2} |0 {\cal i} 
= - \psi^i_{-1/2} |0 {\cal i} \, ,
\label{rule-2}
\ee
for the NN directions, and
\be
\Omega \, \psi^9_{-1/2} |0 {\cal i} = \psi^9_{-1/2} 
|0 {\cal i} \, ,
\label{rule-3}
\ee
for the DD direction.
Thus the $U(1)$ gauge field is projected out and only the
massless transversal scalar survives.
In the $(-1)^F$ NS odd part we have
\be
\Omega |0 {\cal i} = - |0 {\cal i} \, ,
\label{rule-1}
\ee
so the tachyon is projected out, and this D8$_+$ brane is in principle stable.

One can also consider a system of $n$ D8$_+$-branes in type 0.
However, the worldvolume theory will be given by
a non-abelian vector field and a tachyon 
in the adjoint representation of $SO(n)$, and a transversal scalar in the
symmetric representation of $SO(n)$. 
Thus a system of
$n$ D8$_+$ (D8$_-$) branes is in general not stable in type 0 string theory.

We find that a single 
D8$_+$ brane in type 0 is in principle stable, and only contains
a scalar in its worldvolume theory. In order to complete the 
analysis we must consider the open strings which eventually
appear stretching between the D8-brane and the D9-branes.
There are several possibilities that we describe next and which
are summarized in table \ref{table-1}.

\vskip 12pt
\noindent{\bf Stable D8-brane in Type 0 with D9-branes}
\vskip 6pt
 
In type 0 theory with D9-branes in the background there are also 8-9 
open strings,
stretching between the D8$_+$ brane and the D9 and anti-D9 branes. 
Consider the case of a D8$_+$ brane in the background of
32 D9$_+$ branes and 32 anti-D9$_+$ branes. There are then
64 tachyons in the spectrum of the 8-9 strings which appear 
in the NS-sector, for which
\be
L_0 - {3\over 8} = 0
\ee
on the physical states.
On the other hand, if we have 
a D8$_+$ brane in the background of
32 D9$_-$ branes and 32 anti-D9$_-$ branes,
there appear no tachyons in the 8-9 strings since these only contain
a R-sector. Thus we find that a single
D8$_+$ (D8$_-$) brane is stable in type 0 if there are only 
D9$_-$ (D9$_+$) branes in the background. If we disregard the NS-NS
tadpole, we can have any gauge group of the form $SO(n) \times SO(n)$.

Notice that although in type I the D8-brane
is not stable \cite{Lerda-nonBPS}, it can be stable in
type 0. In fact, this can also be derived from the
non-BPS D8-brane of type I.
Consider the construction of type 0 from type I with the orbifold
$(-1)^{F_s} \cdot {\cal I}$ as explained in section 1,
where ${\cal I}$ is taken to be of the form (\ref{cal-I}).
There are tachyons in the NS-sector of 
the 8-9 strings in type I, which can be represented by
\be
| 0 {\cal i} \otimes \Lambda^a \, ,\qquad  a = 1, \dots 32 \, ,
\ee
where $\Lambda^a$ denotes the Chan-Paton factor of the 8-9 
strings\footnote{Since the theory is unoriented this sector contains in fact a 
certain combination of 8-9 and 9-8 strings.}.
The orbifold symmetry has a natural action on these Chan-Paton factors:
\be
| 0 {\cal i} \otimes \Lambda^a \, 
\longrightarrow | 0 {\cal i} \otimes {\cal I}{}^a{}_b \Lambda^b \, ,
\label{cal-I-on-tachyons}
\ee
which can be used to project out $N$ of the tachyons.
This also keeps $32-N$ tachyons, coming from 
the strings stretching between
the D8-brane and $32-N$ D9-branes\footnote{Recall
that these $32-N$ D9-branes are of a different type
from the other $N$ D9-branes.
In fact, this projection also keeps the R-sector only in the strings
stretched between the D8 and the first $N$ D9-branes.}. 
In the twisted sector one has
32 anti-D9-branes, which give rise to $32$ extra tachyons in the 8-9 sector. 
Finally, one performs
again the projection onto invariant states under the transformation
(\ref{cal-I-on-tachyons}). As a result we obtain
$2(32-N)$ tachyons coming from the
D8$_+$-D9$_+$ and D8$_+$-anti-D9$_+$ sectors,
or similarly, from the D8$_-$-D9$_-$ and D8$_-$-anti-D9$_-$ sectors.
In particular, using this procedure we obtain a stable D8-brane in type 0
with $SO(32) \times SO(32)$ 
gauge group for the case ${\cal I} = - \unitmatrixDT_{32}$.
Thus we find again that a non-BPS D8$_+$ (D8$_-$) branes is stable
if there are only D9$_-$ (D9$_+$) branes in the background.

\vskip 12pt
\noindent {\bf Stable D8-brane in Type 0 without D9-branes}
\vskip 6pt

We can make the choice of not introducing
any D9-brane at all in type 0. 
The resulting theory has no open string sector, contains
a tachyon in the closed string sector and a massless NS-NS tadpole.
In this situation there is no 8-9 open string sector, so that a single D8$_+$ 
(D8$_-$) brane is stable. Moreover, since the spectrum of the open strings
stretched between a D8$_+$ and a D8$_-$ is in the 
R-sector (see (\ref{amplitude:D8+/D8-})), we can also consider 
the bound state D8$_\pm$, which is also stable.


\begin{table}[!ht]
\renewcommand{\arraystretch}{1.5}
\begin{center}
\begin{tabular}{|c|c|c|}
\hline
Stable Branes & Type 0 Background & Gauge Symmetry\\
\hline
D7$_+$ \, , D8$_+$ & &\\
D7$_-$ \, , D8$_-$ & No D9-branes & $-$ \\
D7$_\pm$  \, , D8$_\pm$ & & \\
\hline
D7$_+$ \, , D8$_+$ & $n$ D9$_-$ and $n$ anti-D9$_-$  & $SO(n) \times SO(n)$\\ 
D7$_-$ \, , D8$_-$ & $n$ D9$_+$ and $n$ anti-D9$_+$  & $SO(n) \times SO(n)$\\ 
\hline
\end{tabular}
\caption{\small {\bf Stable D7 and D8 branes in type 0.}
In this table we show for which cases a single D7 or D8 brane
can be stable in type 0. The bound states D7$_\pm$ and D8$_\pm$ are also 
considered. We also indicate the gauge group when we have
$n$ D9 anti-D9 pairs, with $n$ arbitrary.}
\label{table-1}
\end{center}  
\end{table}
\renewcommand{\arraystretch}{1}

We can perform a similar analysis for a non-BPS D7-brane in
type 0, using the results of \cite{Lerda-nonBPS}.
We find a stable non-BPS D7$_+$ (D7$_-$) brane
in type 0 for the same cases as for the D8$_+$ (D8$_-$) brane.
We summarize these results in table \ref{table-1}.

\section{Tachyon Condensation in Type 0}

A generic background of type 0 contains an equal number of
D9-branes and anti-D9-branes. This is in fact a natural setting for the
study of tachyon condensation. 
If the D9 and anti-D9 branes annihilate into the vacuum, we are left
with the O9 and the anti-O9 planes, which originate a negative tree-level
cosmological constant \cite{Kachru-Kumar-Silverstein}.
We can consider instead a tachyon condensation for which a D9 anti-D9
pair gives rise to a non-BPS D8-brane. 
In this section we consider this type of condensation
and discuss the stability of the resulting configurations.
We start with the most general case of type 0 with symmetry
$SO(n) \times SO(m) \times SO(n) \times SO(m)$. We also consider other cases
included in this one, which are obtained by letting some of the
D9 anti-D9 pairs to condense into the vacuum.


\vskip 12pt
\noindent {\bf $SO(n) \times SO(m) \times SO(n) \times SO(m)$ symmetry}
\vskip 6pt

Let us consider type 0
with gauge group $SO(n)\times SO(m) \times SO(n) \times SO(m)$, 
originated by $n$ D9$_+$ and
$m$ D9$_-$ branes, and the corresponding antibranes. 
There is moreover an extra $\Z_2$ symmetry corresponding to
the exchange D9$_+$ $\leftrightarrow$ D9$_-$.
In this case there are tachyons
in the $({\bf n},{\bf 1},{\bf n},{\bf 1})$ and
$({\bf 1},{\bf m},{\bf 1},{\bf m})$ representations.
Consider the condensation of all the D9 anti-D9 pairs
of a given type, for instance D9$_-$, into D8-branes. 
We obtain $m$ non-BPS D8$_-$ branes
in the background of $n$ D9$_+$ anti-D9$_+$ pairs.
Although we find no tachyons in the 8-9 strings, there are tachyons
in the 8-8 strings as shown in section 2, so the system 
of $n$ D8-branes is not stable. A similar result is obtained if we let the 
system condense to $m$ non-BPS D8$_-$ branes plus $n$
non-BPS D8$_+$ branes.

	We can compare this result with the analysis of the
symmetries of the tachyon potential, in a similar fashion as in
\cite{Schwarz-talk}. The tachyon potential 
has a $SO(n) \times SO(m) \times SO(n) \times SO(m)$ symmetry, 
hence the vacuum manifold,
$S^{n-1} \times S^{m-1} \times S^{n-1} \times S^{m-1}$, is connected. 
Thus there are no stable kink solutions
to this potential. The system is then expected to decay into the vacuum.


\vskip 12pt
\noindent {\bf $SO(n) \times SO(n) \times \Z_2$ symmetry}
\vskip 6pt

In the above configuration, 
we can consider the condensation of $k$, $k<m$,
D9$_-$ anti-D9$_-$ pairs
into D8$_-$ branes and $m-k$ pairs into the vacuum.
The result is a set of $k$ non-BPS D8$_-$ branes in the background of
$n$ D9$_+$ anti-D9$_+$ pairs. 
For $k>1$ the D8-branes are not stable since there are tachyons in the
8-8 sector. Moreover, the background is not stable since there are tachyons
in the 9-9 sector.
If the $n$ D9$_+$ anti-D9$_+$ pairs decay into the vacuum, the result is then
a set of $k$ non-BPS D8$_-$ branes in type 0 without D9-branes. According 
to the results of section 2 this is only stable for $k=1$.

	We compare now with the analysis of the symmetries of 
the tachyon potential. The condensation of $k$ 
D9$_-$ anti-D9$_-$ pairs into D8$_-$ branes and $m-k$ pairs 
into the vacuum requires that some of the components
of the $({\bf 1},{\bf m},{\bf 1},{\bf m})$ tachyon must
condense independently.
In order for this to be possible
the group associated to the D9$_-$ branes,
the symmetry $SO(m) \times SO(m)$ must be
broken down to $SO(m-k) \times SO(m-k) \times SO(k)
\times SO(k)$. The part $SO(m-k) \times SO(m-k)$
is associated to the D9-branes that condense into the vacuum, so we need not
to consider it any longer.
The relevant symmetry 
group in order to seek for kink solutions is 
$SO(n) \times SO(k) \times SO(n) \times SO(k)$, originated by
$k$ D9$_-$ anti-D9$_-$ pairs , which we plan to condense into D8$_-$ branes;
and by $n$ D9$_+$ anti-D9$_+$ pairs that we leave untouched in the background.
For $k>1$ the vacuum manifold associated to the tachyon potential 
is again connected and we find no stable kinks.

	If $k=1$, we have $n$ D9$_+$ anti-D9$_+$ pairs 
and just one D9$_-$ anti-D9$_-$ pair\footnote{We do not consider 
the rest of the D9$_-$ anti-D9$_-$ pairs which 
have condensed into the vacuum.}. 
The symmetry of the tachyon potential
is $SO(n) \times SO(n) \times \Z_2$ and
the vacuum manifold corresponds to two 
discrete copies of $S^{n-1} \times S^{n-1}$. This is 
not connected and accepts a topological 
stable kink solution corresponding\footnote{One can also consider the
condensation of a higher tachyonic mode of a single
D9 anti-D9 pair. This will give rise in general to 
$n+1$ kinks and $n$ ($n+1$) anti-kinks, which will correspond to
$2n+1$ ($2n+2$) non-BPS D8-branes. I thank A.~Sen for drawing my attention
to this possibility.} to
a stable non-BPS D8$_-$ brane in the background of
$n$ D9$_+$ anti-D9$_+$ pairs. These $n$ pairs are not stable
due to the tachyons in the 9-9 sector, so that they can decay into the vacuum 
and we are left with a single stable non-BPS D8$_-$.


\vskip 12pt
\noindent {\bf $SO(n) \times SO(n)$ symmetry}
\vskip 6pt

If all the D9 anti-D9 pairs of a given type decay into the vacuum, we are left
with type 0 with a gauge group
of the form $SO(n) \times SO(n)$. This symmetry is originated by
$n$ D9$_+$ (or D9$_-$) branes and the corresponding $n$ antibranes.
This theory has a tachyon in the $({\bf n},{\bf n})$ representation.
The analysis of this case is similar to the one above.
	
If we consider the condensation of
$k$ pairs, $n \geq k >1$, we obtain $k$ non-BPS D8$_+$ (D8$_-$) 
branes in the background
of $n-k$ D9$_+$ anti-D9$_+$ (D9$_-$ anti-D9$_-$) pairs.
This is not stable since there are tachyons
in the 8-8 strings. If the $n-k$ D9 anti-D9 pairs 
do not decay into the vacuum, there are moreover tachyons in the
8-9 and 9-9 strings.
The relevant tachyon potential has $SO(k) \times SO(k)$ symmetry,
so that the minimum describes a vacuum manifold of the form
$S^{k-1} \times S^{k-1}$. This manifold is connected so there is no topological
stable kink. The case $k=1$ is included in the discussion below.

\vskip 12pt
\noindent {\bf $\Z_2 \times \Z_2$ symmetry}
\vskip 6pt

A special case is when we let all the D9 anti-D9 pairs 
condense into the vacuum except for a pair of each type,
i.e. we are left with one D9$_+$ anti-D9$_+$ pair and one D9$_-$ anti-D9$_-$ 
pair. 
The symmetry $SO(n) \times SO(m) \times SO(n) \times SO(m)$ is then broken
down to $\Z_2 \times \Z_2$.
We have two tachyon fields with a
$\Z_2$ symmetry each. Moreover, there is one
extra $\Z_2$ symmetry corresponding 
to the exchange D9$_+$ $\leftrightarrow$ D9$_-$. 
Let us analyze the possible kink solutions in this potential.
One possibility is that one tachyon condenses into a kink solution and the 
other condenses into the vacuum. This gives us two possible kinks,
which are indistinguishable because of the extra $\Z_2$ symmetry.
The result is a stable non-BPS D8 brane, either D8$_+$ or D8$_-$, in type 0 
without D9-branes. One can of course leave one of the D9 anti-D9 pairs
intact. The symmetry of the potential is then $\Z_2$ and we obtain,
after tachyon condensation, a non-BPS D8$_+$ (D8$_-$) brane
in type 0 with a D9$_-$ anti-D9$_-$ (D9$_+$ anti-D9$_+$)
pair in the background, which is also stable.

Another possibility is when both tachyons condense into a kink.
Notice that both tachyon fields have dependence on the same compact direction
$X$ along the D9-branes. 
The position of the zero of the first tachyon kink along the $X$-axis 
is free to be chosen by reparametrisation invariance. However, the
relative position of the zero of the second tachyon is not.
Consequently we obtain a 1-parameter family of kink-pairs, where the parameter 
indicates the relative distance between the zeroes of the two kinks.
On the other hand, the $X$-axis becomes 
the direction transverse to the non-BPS
D8-branes after tachyon condensation \cite{Sen-2}. Thus the distance-parameter
indicates the relative distance between the non-BPS D8$_+$ and the non-BPS
D8$_-$ after tachyon condensation. This means that in the procedure of tachyon
condensation we obtain a stable non-BPS D8$_+$ - D8$_-$ pair 
with a relative separation distance. 
In particular, when the zeros of both kinks
coincide, we obtain the stable bound state D8$_\pm$.


\section{Conclusions}

We have found stable non-BPS D7 and D8 branes in type 0.
We have made use of the fact that the background of D9-branes in type 0
is not completely determined by the tadpole cancellation, and that
we have the freedom of choosing between two types of D9-branes. Moreover,
disregarding the NS-NS tadpole, which can be removed by other means,
we can have any number of D9 anti-D9 pairs in the background. 
The results are summarized in table \ref{table-1}.

The open string tachyon in the
D9 anti-D9 pair can condense to give rise to the non-BPS D8-brane
described in this paper.
We have analyzed the tachyon condensation in type 0 for 
several generic configurations of D9-branes, and found that
only when there is a $\Z_2$ symmetry involved
the condensation yields a stable configuration. This corresponds
to having only one D9 anti-D9 pair of at least one type,
either D9$_+$ or D9$_-$.
For the particular case of the $\Z_2 \times \Z_2$ symmetry there is also
a freedom to separate the resulting non-BPS branes in the
procedure of tachyon condensation.
 
We notice that in type 0 the NS-NS tadpole can be eliminated
either by a Fischler-Susskind mechanism, or by adding
D9 and anti-D9 branes. The Fischler-Susskind mechanism
introduces a spacetime-dependent metric and dilaton.
On the other hand, the D9 anti-D9 pairs can give rise to non-BPS
D8-branes after tachyon condensation. In general, a background with 
D8-branes has associated a metric and a dilaton with non-trivial 
dependence on the spacetime \cite{Polchinski-Witten}. It would be
interesting to investigate further the background of type 0 in the presence 
of these non-BPS D8 branes and see whether there is any relation
with a Fischler-Susskind mechanism in type 0.

Finally, type 0 with gauge group $SO(32) \times SO(32)$
was conjectured to be S-dual to the compactification of the 26-dimensional
bosonic string on the $SO(32)$ weight lattice 
\cite{Bergman-Gaberdiel-1}. 
It would be interesting to find out whether
this duality can be extended to more generic type 0 backgrounds, like those 
described in this paper. 
Moreover, it would be very interesting to see to which 
states correspond the stable non-BPS D7 and D8 branes in the
dual theory.

\section*{Acknowledgements}

I would like to thank A.~Sen for helpful discussions and 
for his comments on a draft of the paper.
I am also happy to thank  A.~Ach{\'u}carro, 
A.~Lerda and J.L.~Ma{\~ n}es for useful discussions,
and especially to R.~Argurio and D.~Matalliotakis
for their comments on the paper and encouragement. I am also thankful to the 
organizers of the {\it Advanced School on Supersymmetry in the Theories of 
Fields, Strings and Branes} in Santiago de Compostela, where part
of this work was carried out, for providing a great 
physics and non-physics environment.

\end{document}